\documentclass[%article,
reprint,
amsmath,
amssymb,
prb,
floatfix]{revtex4-2}

\usepackage{comment}
\usepackage[T1]{fontenc}
\usepackage{xcolor}
\usepackage[outline]{contour}
\usepackage{graphicx}
\usepackage{bm}
\usepackage[f]{esvect}
\usepackage{siunitx}
\usepackage{mathtools}
\usepackage{physics}
\usepackage[allcolors=blue,colorlinks]{hyperref} %draft=false,
\usepackage[english]{babel}

\newcommand{\pp}{\mathcal{P}}
\newcommand{\lio}{\mathcal{L}}
\newcommand{\qqq}{\mathcal{Q}}
\renewcommand{\Trace}[2]{ {\rm Tr}_{\rm #1}\left\{ #2 \right \} }
\newcommand*\diff{\mathop{}\!\mathrm{d}}

\contourlength{0.2pt}
\contournumber{10}

\newcommand{\tvec}{\vec{T} }

\newcommand{\et}{\vec{\rm{e}}_T}

\newcommand{\ey}{\vec{\rm{e}}_y}
\newcommand{\ez}{\vec{\rm{e}}_z}
\newcommand{\eperp}{\vec{\rm{e}}_\perp}

\newcommand{\gm}{\vec{\gamma}^-}

\newcommand{\vecnt}{\vec{n}_T^l}

\begin{document}

\title{Precession of entangled spin and pseudospin in double quantum dots}
\author{Christoph Rohrmeier}
\email{christoph.rohrmeier@ur.de}
\author{Andrea Donarini}
\affiliation{Institute of Theoretical Physics, University of Regensburg, 93053 Regensburg, Germany}

\date{\today}

\begin{abstract} 

Quantum dot spin valves are characterized by exchange fields which induce spin precession and generate current spin resonances even in absence of spin splitting. Analogous effects have been studied in double quantum dots, in which the orbital degree of freedom, the pseudospin, replaces the spin in the valve configuration. We generalize, now, this setup to allow for arbitrary spin and orbital polarization of the leads, thus obtaining an even richer variety of current resonances, stemming from the precession dynamics of entangled spin and pseudospin.  We observe for both vectors a delicate interplay of decoherence, pumping and precession which can only be understood by also considering the dynamics of the spin-pseudospin correlators. The numerical results are obtained in the framework of a generalized master equation within the cotunneling approximation and are complemented by the analytics of a coherent sequential tunneling model. 

\end{abstract}

\maketitle

%%%%%%%%%%%%%%%%%%%%%%%%%%%%%%%%%%%%%%%%%%%
\section{Introduction}

The coherent manipulation of spins is at the heart of quantum information technology. In this regard, \emph{electron spin resonance} (ESR) is a compelling tool able to accomplish for this task in a controlled way \cite{Poole1996,Willke2021}. The basic working principle of ESR requires a constant magnetic field splitting the spin energy levels together with a magnetic field oscillating at the resonant frequency. In a system comprising several spin centers, however, it is very challenging to produce localized magnetic fields which address one spin at a time. To overcome this obstacle, it was shown that, by mixing charge and spin degrees of freedom, it is possible to manipulate the electronic spins also by electrical gates \cite{PioroLadriere2008}. In this particular setup, a Pauli-spin blockade was lifted in a double quantum dot (DQD) - addressing the dots individually - via the non-uniform Zeeman field of an electrically tuned micromagnet.

Triggering spin precession with electrical signals is also achieved without a time dependent driving by pushing the magnetic component into the leads. The \emph{spin resonance without spin splitting} \cite{Hell2015} is obtained, for example, in a quantum dot spin valve. Here, the spin of the quantum dot precesses around the constant exchange field arising by charge fluctuations towards ferromagnetic leads with non-collinear polarization. The spin precession lifts the blockade due to the valve configuration of the leads and a resonant current flows. 
As the resonance condition does not depend on the local magnetic field gradients but rather on the tunneling couplings, scalability of such spin resonant devices is envisaged by means of local gating of the individual dots. 

Exchange fields, generated by the electronic fluctuations, characterize, in general, the dynamics of degenerate, interacting quantum system weakly coupled to the leads \cite{Braun2004, Darau2009, Donarini2010, Karlstroem2011, Hell2013}. Those electronic fluctuations lead to a renormalization of system energies \cite{Wunsch2005,Wetzels2005} which - under the right circumstances - also drive precession dynamics of the system degrees of freedom, similarly to external magnetic fields \cite{Maurer2020}. 

A profuse interest has been attracted by interacting spin valves with increasing complexity. Spin and Coulomb spectroscopy has been achieved with artificial molecules \cite{PioroLadriere2002}, and also pure spin currents have been predicted by pumping protocols in a DQD valve\cite{Riwar2010}. More recently, a continuously electrically tunable quantum dot spin polarization of 80\% \cite{Bordoloi2020} has been realized with individual split gates, and a gate tunable enhanced magnetoresistance has been predicted by exploiting exchange renormalization of the dot levels \cite{Tulewitz2021}.  

The interplay of ferromagnetism and Kondo resonance \cite{Pasupathy2004}  as well as the spatial resolution of spin states \cite{Iacovita2008} has been investigated in single molecule junctions, while  gate field controlled  magnetoresistance has been reported in carbon nanotubes \cite{Sahoo2005, Dirnaichner2015}. 

In a recent publication \cite{Rohrmeier2021}, we extended the concept of exchange field mediated spin resonances to include the pseudospin. There, the orbital degree of freedom of a DQD yields distinctive resonances in transport setups where a geometrically induced pseudospin valve suppresses the current. Interestingly, the resonance is split in presence of parallel polarized ferromagnetic leads, thus highlighting the emergence of a tunable synthetic spin-orbit coupling. 

Moreover, an easier tunability of the system parameters characterizes, in general, the pseudospin degree of freedom. The polarization of the leads, for example, is for the pseudospin a property of the interface, and as such tunable in strength and direction,  together with the tunneling amplitudes. The spin polarization, however, relies on material properties, hardly tunable and, above all, difficult to integrate, for example, in semiconductor heterostructures.     

In this work, we extend the analysis to an interacting DQD spin valve with generic spin and pseudospin polarizations. The interplay of spin and pseudospin produces, within the one-particle Coulomb diamond, a rich pattern of current resonances, strongly modulated by the direction and strength of the pseudospin polarization. Moreover, Coulomb interaction induces pseudospin anisotropy on the DQD, thus defining a pseudospin hard axis. The angle between this axis and the pseudospin polarization direction of the leads is the crucial parameter tuning the entanglement between spin and pseudospin on the DQD.  

The paper is organized as follows: In Sec.~\ref{sec:model}, we introduce our model of a DQD spin valve. We elaborate in Sec.~\ref{sec:transport_theory} on the transport theory necessary for a consistent description of the expected interference effects. Numerical results of transport characteristics are presented in Sec.~\ref{sec:numerical_results}. The analytics of a coherent sequential tunneling model follow in Sec.~\ref{sec:CST_model}, where we analyze and generalize the appearing current resonances. In Sec.~\ref{sec:limiting_cases} special relevance is given to important limiting cases of our setup, while Sec.~\ref{sec:entanglement} is dedicated to the emerging entanglement between spin and pseudospin. The discussion of feasible experimental implementation of our concept can be found in Sec.~\ref{sec:experimental_realization} and it is followed by an overall conclusion (Sec.~\ref{sec:conclusion}).

%%%%%%%%%%%%%%%%%%%%%%%%%%%%%%%%%%%%%%%%%%%%
\section{Model}
%%%%%%%%%%%%%%%%%%%%%%%%%%%%%%%%%%%%%%%%%%%%
\label{sec:model}
We consider a spinfull DQD weakly coupled in parallel to ferromagnetic leads. The setup is described by the following system-bath Hamiltonian,
\begin{equation}
    \label{eq:hamiltonian}
    \hat{H} = \hat{H}_\text{DQD} +\hat{H}_\text{leads} + \hat{H}_\text{tun},
\end{equation}
which contains contributions of the DQD, the leads and the tunneling coupling. The Hamiltonian of the DQD reads
\begin{equation}
    \label{eq:H_DQD}
    \begin{split}
        \hat{H}_\text{DQD} = \sum_{i =1,2} \left[\left(\varepsilon_0 +  \si{e} V_\text{g} +  \right)\hat{n}_{i}
        +U \hat{n}_{i\uparrow} \hat{n}_{i\downarrow}\right]+V \hat{n}_{1} \hat{n}_{2},
    \end{split}
\end{equation}
where $\hat{n}_{i\sigma}= \hat{d}_{i\sigma}^\dagger \hat{d}_{i\sigma}$ is the number operator for electrons on the $i$th dot with spin $\sigma$, $\hat{d}_{i\sigma}$ being the corresponding electronic annihilation operator, $\si{e}$ the electronic charge, and $\hat{n}_i = \sum_\sigma \hat{n}_{i\sigma}$. The gate voltage is denoted by $V_\text{g}$ and the on-site energy by $\varepsilon_0$. Moreover, we differentiate between $U$, the local and $V$ the interdot Coulomb interaction, with the general condition $U > V$ favoring electron delocalization over the full DQD.
The electrons in the DQD are completely characterized by a spin and an orbital degree of freedom. The latter identifies the occupation of a specific dot, simply labelled as "1" or "2". In analogy to the spin degree of freedom, we introduce, for the orbital one, a pseudospin description. The three components of the pseudospin operator are given by
\begin{equation}
    \label{eq:Pseudospin}
    \hat{T}_\alpha = \frac{1}{2}\sum_{\tau i j}  \hat{d}^\dagger_{i\tau} \sigma^\alpha_{i j}  \hat{d}_{j\tau},
\end{equation}
where $\alpha = x,y,z$ and $\sigma^\alpha$ are the Pauli matrices. The occupation numbers for dot "1" and dot "2" can thus be expressed as
\begin{equation}
    \hat{n}_{1,2}= \frac{\hat{N}}{2}\pm \hat{T}_z,
\end{equation}
where $\hat{N} = \hat{n}_{1} +\hat{n}_{2}$ is the total particle number operator of the system. 
A finite component of the pseudospin in the $z$ direction is thus associated to an excess occupation of the dot "1" with respect to the dot "2". $H_\text{DQD}$ can be reformulated in terms of the pseudospin operators as
\begin{equation}
    \label{eq:H_DQD_pseudo}
    \hat{H}_\text{DQD} = \left(\varepsilon -\frac{U}{2}\right)\hat{N}+\frac{U+V}{4}\hat{N}^2+\left(U-V\right)\hat{T}_z^2,
\end{equation}
where $\varepsilon=\si{e} V_\text{g} + \varepsilon_0$. 

The eigenstates of $\hat{H}_\text{DQD}$ can be classified in terms of spin and pseudospin quantum numbers. The latter are crucial for the understanding of the DQD dynamics considered in this manuscript. We thus analyze them with considerable detail. The extreme occupation numbers (0 and 4) are both spin as well as pseudospin singlets. 
The one-particle sector, instead, is spanned by four degenerate states. On one hand, we can identify them as $\ket{\sigma,0}$ or $\ket{0,\sigma}$, i.e. as states with an electron with spin $\sigma$ occupying respectively the first or the second dot. Alternatively, according to Eq.~\eqref{eq:Pseudospin}, these states of spin $S_z = \pm 1/2$ are also characterized by a pseudospin $T_z= \pm 1/2$. A state with a finite pseudospin component in the $x$ or $y$ direction, requires, instead, a coherent superposition of localized states. Analogous considerations concern the three-particle sector, understood in terms of states with a single hole.

\begin{figure}
    \includegraphics[width=0.7\columnwidth,draft=false]{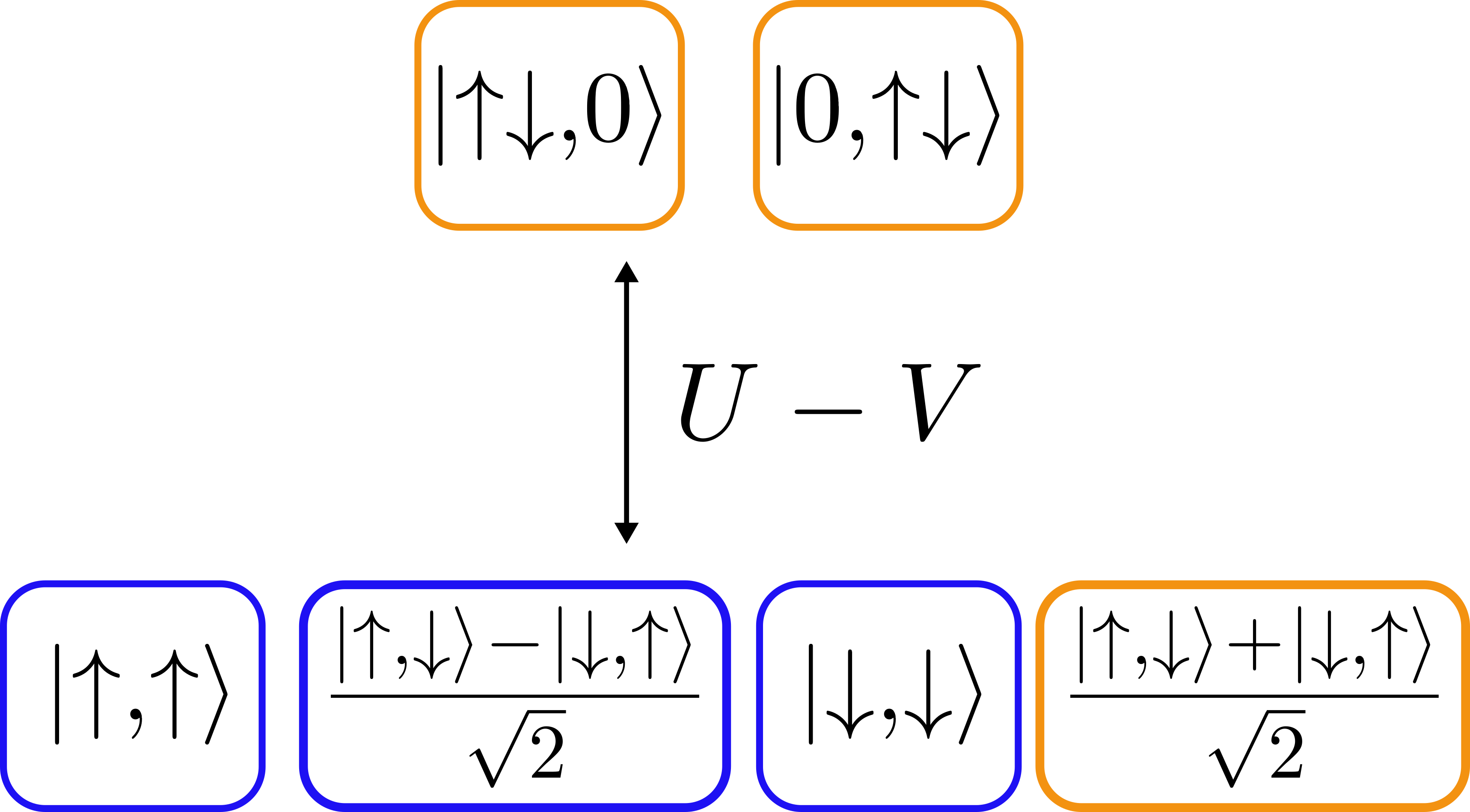} 
    \caption{\label{fig:energy_landscape} Energy splitting of the two-particle states: The spin-triplet, pseudospin-singlet states are depicted in blue ($S_z=0,\pm 1$ and $T=0$). The pseudospin anisotropy splits the pseudospin-triplet, spin-singlet states (highlighted in orange; $T_z=0,\pm 1$ and $S=0$) energetically.}
\end{figure}

The perfect symmetry between spin and pseudospin observed in the sectors with the outer occupation numbers is broken in the two-particle sector. The latter highlights, instead, against its perfect spin isotropy a pseudospin anisotropy of the DQD. As $U > V$, the delocalization of the two electrons is energetically favored and the pseudospin develops an easy  $x$-$y$ plane as indicated by the last term of Eq.~\eqref{eq:H_DQD_pseudo}. The latter vanishes in the zero- and four-particle subspaces (both pseudospin singlets) while it reduces to a constant energy shift when evaluated on the one- and the three-particle subspaces (corresponding both to pseudospin doublets). In the two-particle subspace, we have to deal with a combination of spin/pseudospin triplets and singlets. An overview of the resulting six states is visualized in Fig.~\ref{fig:energy_landscape}. The fourfold degenerate groundstate consists of the three spin-triplet, pseudospin-singlet states (blue) and the $(T_z=0)$-pseudospin-triplet, spin-singlet state (orange). The remaining $(T_z = \pm 1)$-pseudospin-triplet states are split off with an energy $U-V$ higher than the one of the ground state. This energy splitting in the pseudospin space will be the key to understand the richness of the observed current resonances.

The leads are baths of (effectively) non-interacting fermions, described by the Hamiltonian ${H}_\text{leads}=\sum_{b } \varepsilon_{b} \hat{c}_{b}^\dagger \hat{c}_{b}$. 
The collective index $b=\lbrace l\vec{k}\sigma_l \rbrace$ labels an electron according to its location ($l = {\rm L/R}$ for the left/right lead), momentum $\vec{k}$ and spin $\sigma_l$ along the spin quantization axis of the $l$-lead. The quantization axes of the system and of the individual leads do not necessarily coincide, and a more general choice helps in the description of non-collinearly polarized leads. The annihilation operator $\hat{c}_{b}$ destroys the lead electron with the corresponding energy $\varepsilon_{b}$. Furthermore, we assume the leads to be spin polarized with a spin polarization for the lead $l$ defined as $P^l_{S} = (g_{l\uparrow_l} - g_{l\uparrow_l})/(g_{l\uparrow_l} + g_{l\uparrow_l})$, being $g_{\ell\sigma}$ the spin resolved density of states for the lead $l$.  We keep the leads at the same temperature $T$, with the electrochemical potentials $\mu_l$ modulated by the external bias $\mu_{\rm L,R} = \pm \si{e}V_{\rm b}/2$.

Finally, the tunneling Hamiltonian reads
\begin{equation}
    \hat{H}_\text{tun} = \sum_{b i \sigma} t_{b,i\sigma}\, \hat{c}_{b}^\dagger\,\hat{d}_{i\sigma} +t_{b,i\sigma}^*\,\hat{d}^\dagger_{i\sigma}\,\hat{c}_{b},
\end{equation}
where the tunneling amplitudes $t_{b,i \sigma}$ connect the operators of the leads with the four system operators $\{\hat{d}_{1\uparrow},\hat{d}_{1\downarrow},\hat{d}_{2\uparrow},\hat{d}_{2\downarrow}\}$. The tunneling amplitudes allow us to define the tunneling rate matrix
\begin{equation}
    \label{eq:Gamma_def}
    \Gamma^l_{i\sigma,j\sigma'}(E) = \frac{2\pi}{\hbar} \sum_{\vec{k}\sigma_l} t_{l\vec{k}\sigma_l ,i \sigma}^{\ast}\,t_{l\vec{k}\sigma_l ,j \sigma'}\, \delta(E-\varepsilon_{l\vec{k}\sigma_l }),
\end{equation}
which incorporates the properties of the tunneling process. In this work, we assume negligible intrinsic spin-orbit interaction and very localized dot wave functions, so that we can factorize the tunneling rate matrices into an orbital/pseudospin and a spin component. We write thus, for the tunneling rate matrix for the $l$-lead
\begin{equation}
 \label{eq:Gamma_matrix}
    \Gamma^l = \Gamma^l_0 \left(\frac{\openone_2}{2} + \frac{P_T^l}{2} \vec{n}_{T}^l \cdot \vec{\sigma}\right) \otimes \left(\frac{\openone_2}{2} +\frac{P_S^l}{2} \vec{n}_{S}^l \cdot \vec{\sigma}\right),
\end{equation}
where we introduced the bare tunneling rate $\Gamma^l_0$, the spin/pseudospin polarization strength $P_{S/T}^l$ and direction $\vec{n}_{S/T}^l$. The pseudospin polarization of the lead allows for a simple physical interpretation, in connection to the pseudospin formulation of the system Hamiltonian. Full pseudospin polarization in the $z$-direction indicates an exclusive coupling to the dot "1" or, depending on the direction, to the dot "2". Components in the $x$-$y$ plane describe instead coherent tunneling to both orbitals. 

\begin{figure}
    \includegraphics[width=\columnwidth,draft=false]{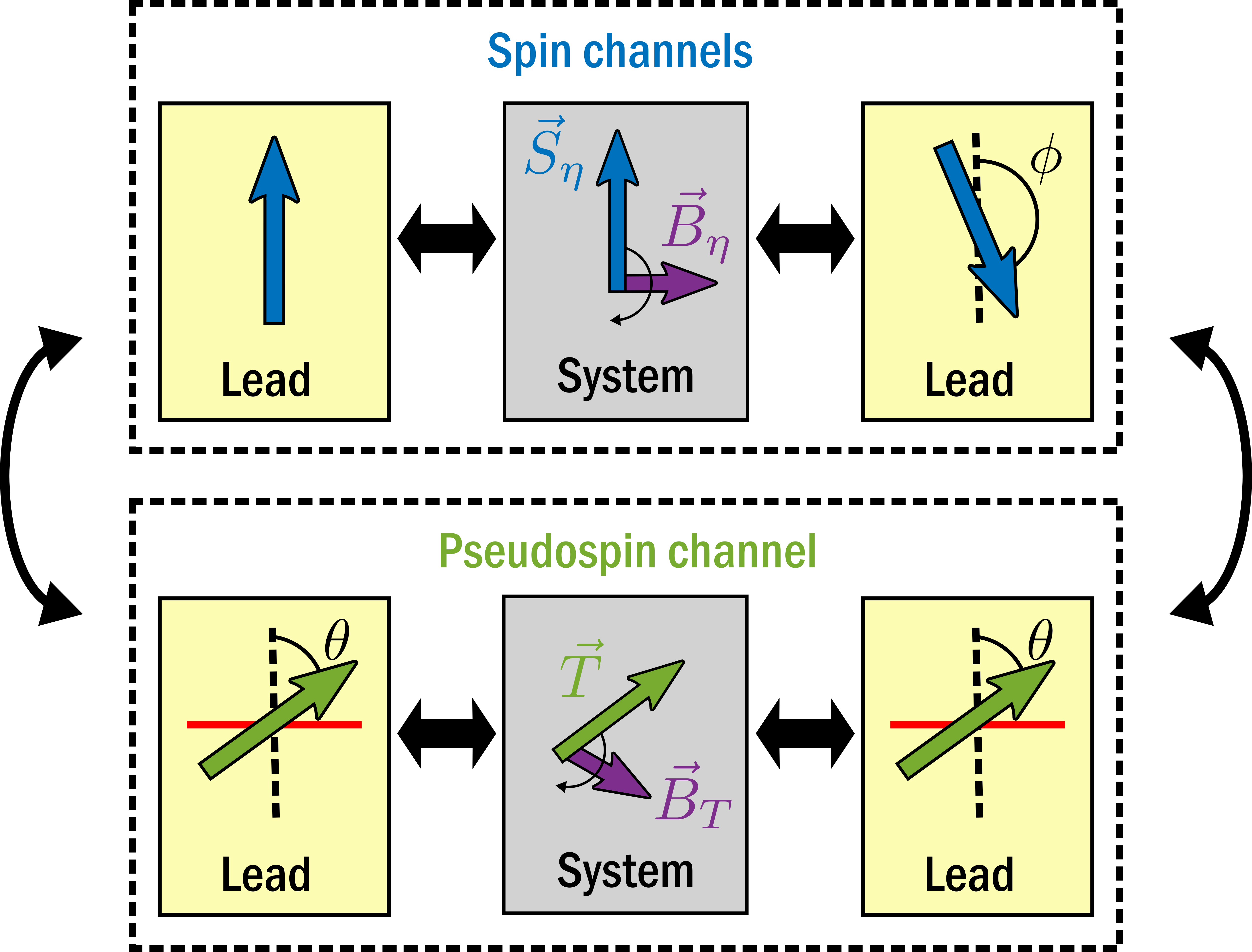}
    \caption{\label{fig:setup2}  Interplay of spin and pseudospin channels determines the transport through the system: In the spin space, the polarization vectors of the leads are almost antiparallel ($\phi\approx\pi$) which translates into a spin valve configuration. Through pseudo exchange fields (purple) one can rotate the spin of the system, thus lift the current suppression of the spin valve. In the pseudospin space, we consider parallel polarization of the leads. Since there is a preferential plane in the pseudospin, indicated by the red lines, one defines the polarization direction of the leads in respect with this plane ($\theta$). The pseudospin of the system can precess under the influence of the pseudo exchange field $\vec{B}_T$.}
\end{figure}

The tuning of the system geometry yields different tunneling setups with different tunneling rate matrices. In Fig.~\ref{fig:setup2} one of the tunneling configuration considered in this manuscript is visualized. We distinguish for clarity between spin and pseudospin channels, even if, except of some limiting cases, the full system dynamics results from their interplay, as suggested by the curved arrows. In the spin degree of freedom, we will limit our considerations to a valve configuration, where the polarization vectors of the leads are almost antiparallel. The latter leads to an overall suppression of the current through the tunneling junction associated to a spin accumulation. The pseudospin polarization vectors of the leads are, instead,  parallel to each other, though they do not coincide, in general, with the pseudospin hard axis of the DQD (indicated by dashed black line) nor do they belong to the easy plane (indicated schematically by the solid red lines). 

Due to the spin isotropy of the DQD and its rotational invariance around its hard axis, we can parametrize the polarization vectors with just two angles:
\begin{align}
\label{eq:pseudospin_angle}
    &\vec{n}_{S}^{\rm L}=\left(0,0,1\right),\vec{n}_{S}^{\rm R}=\left(\sin{\phi},0,\cos{\phi}\right), \\
    &\vec{n}_{T}^{\rm L/R}=\left(\sin{\theta},0,\cos{\theta}\right)=\vec{n}_{T}.
\end{align}
Moreover, throughout this work we will use equal spin and orbital polarization for the leads ($P_S=P_S^\text{L/R}$; $P_T=P_T^\text{L/R}$).

%%%%%%%%%%%%%%%%%%%%%%%%%%%%%%%%%%%%%%%%%%
%%%%%%%%%%%%%%%%%%%%%%%%%%%%%%%%%%%%%%%%%%

%%%%%%%%%%%%%%%%%%%%%%%%%%%%%%%%%%%%%%%%%%
%%%%%%%%%%%%%%%%%%%%%%%%%%%%%%%%%%%%%%%%%%

%%%%%%%%%%%%%%%%%%%%%%%%%%%%%%%%%%%%%%%%%%
\section{Transport theory}
%%%%%%%%%%%%%%%%%%%%%%%%%%%%%%%%%%%%%%%%%%
\label{sec:transport_theory}
We investigate the transport characteristics of the DQD spin valve via a generalized master equation (GME) for the reduced density matrix.  In particular, we consider a perturbative approach in the tunneling coupling to the leads. We complement the lowest perturbative order --the sequential tunneling approximation-- with the cotunneling contributions, in which also the simultaneous and correlated two-electron tunneling events are taken into account. 
Most of the current features presented in this manuscript already appear in sequential tunneling. Thus, the cotunneling approximation also serve to demonstrate the robustness of the observed effects. 

By numerical integration of the GME, we evaluate the steady state reduced density matrix, from which observables like the current, the spin, the pseudospin, the populations of the dots and the concurrence are evaluated. We will adhere very closely to the formalism presented in our recent publication \cite{Rohrmeier2021} and based on the Nakajima-Zwanzig projection operator technique \cite{Nakajima1958,Zwanzig1960}. The equation of motion for the projected component of the density operator $\mathcal{P}\hat{\rho}$ reads \cite{Breuer2002}
\begin{equation} 
    \label{eq:nakajimazwanzig}
    \mathcal{P}\dot{\hat{\rho}}(t) = \mathcal{L}_{\text{DQD}}\mathcal{P}\hat{\rho}(t)+ \int_{0}^{t} \diff s \; \mathcal{K}(t-s) \pp \hat{\rho}(s),
\end{equation}
where the projector $\mathcal{P} =: \Trace{\rm leads}{\bullet} \otimes \hat{\rho}_{\rm \rm leads}$ is defined as the partial trace over the leads times the equilibrium density matrix of the leads. Furthermore, we introduced the Liouville operator for the DQD: $\mathcal{L}_{\text{DQD}}=:-i/\hbar\big[\hat{H}_\text{DQD},\hat{\rho}\big]$. Analogously defined Liouvilleans corresponding to the other components of the Hamiltonian in Eq.~\eqref{eq:hamiltonian} can be found in the kernel superoperator
\begin{equation}
    \label{eq:kernel_full}
    \mathcal{K}(t)= \pp \lio_\text{tun} \bar{\mathcal{G}}_\qqq (t) \lio_\text{tun} \pp,
\end{equation}
which combines tunneling Liouvillans with the propagator
\begin{equation}
    \label{eq:G_Q}
    \bar{\mathcal{G}}_\qqq (t)= e^{\left( \lio_{\text{DQD}}+\lio_\text{leads}+\qqq \lio_\text{tun} \qqq \right) t}.
\end{equation}
containing, moreover, the complementary projector $\qqq = 1 - \mathcal{P}$. 

Equation~\eqref{eq:nakajimazwanzig} is exact and it contains all orders in the tunneling Liouvillian $\mathcal{L}_{\rm tun }$ as it is easily verified by inspection of Eq.~\eqref{eq:G_Q}. Moreover, Eq.~\eqref{eq:nakajimazwanzig} also captures memory effects, as the dynamics of the reduced density matrix at time $t$ depends on the state of the system at all previous times. In the present work, we concentrate, though, only on the steady state of the system, defined as $\hat{\rho}^\infty_{\text{red}} \coloneqq \text{Tr}_\text{leads}\left\lbrace \hat{\rho}(t\rightarrow \infty) \right\rbrace$. With the help of the Laplace transformation, the convolutive form of the kernel, and the final value theorem, we obtain the following equation for the stationary reduced density operator \cite{Koller2010_1,Leijnse2010,Koller2010_2,Niklas2018}:
\begin{equation}
    \Trace{\rm leads}{\left( \lio_{\text{DQD}}+\mathcal{\tilde{K}} \right) \hat{\rho}^{\infty}_{\text{red}}\otimes \hat{\rho}_{\rm leads}} = 0
\end{equation}
with
\begin{equation}
    \label{eq:kernel}
    \tilde{\mathcal{K}} = \pp \lio_\text{tun} \sum_{n=0}^\infty \left(\tilde{\mathcal{G}}_0 \qqq \lio_\text{tun} \qqq \right)^{2n}
    \tilde{\mathcal{G}}_0 \lio_\text{tun} \pp,
\end{equation}
where
\begin{equation}
    \tilde{\mathcal{G}}_0 = \lim_{\lambda \rightarrow 0^+} \frac{1}{\lambda - \lio_{\text{DQD}} - \lio_\text{leads}}
\end{equation}
is the Laplace transform of the free propagator for the DQD and the leads, but in the absence of tunneling coupling. Notice that memory effects do not affect the steady state properties of the system and the Markovian limit of the GME would yield the same results. 

For sufficiently small coupling to the leads ($\hbar\Gamma_0 \ll U, k_{\rm B}T$) a perturbative expansion of the propagation kernel in Eq.~\eqref{eq:kernel} in the tunneling Liouvillian is justified. Throughout this work, we will focus on the first two terms in this expansion, namely the sequential tunneling ($n=0$) and the cotunneling terms ($n=1$). We refer to \cite{Koller2010_1,Leijnse2010,Koller2010_2,Niklas2018,Rohrmeier2021} for a more detailed discussion about the sequential and cotunneling Kernels of Eq.~\eqref{eq:kernel} with the evaluation of the respective energy integrals. Any stationary expectation value of a system observable can be obtained as $O = \text{Tr}_{\rm DQD}\{\hat{O} \hat{\rho}_{\rm red}^\infty\}$. 
From the stationary density matrix $\hat{\rho}_{\rm red}^\infty$, also the stationary current at lead $l$ is evaluated,
\begin{equation}
    I_l = \Trace{\rm DQD+leads}{\mathcal{K}_{I_l} \hat{\rho}_{\rm red}^\infty \otimes \hat{\rho}_{\rm leads}},
\end{equation}
with the current kernel $\mathcal{K}_{I_l}$ obtained from the propagator kernel in Eq.~\eqref{eq:kernel} by changing the leftmost tunneling Liouvillian with the current operator,
\begin{equation}
    \hat{I}_l = \frac{i\si{e}}{\hbar}\sum_{{\bm k}\sigma_l a \sigma} 
    t_{l{\bm k}\sigma_l, a\sigma}\,\hat{c}^\dagger_{l{\bm k}\sigma_l}\,\hat{d}_{a\sigma}-
    t^*_{l{\bm k}\sigma_l, a\sigma}\,\hat{d}^\dagger_{a\sigma}\,\hat{c}_{l{\bm k}\sigma_l},
\end{equation}
where $\si{e}$ is the electronic charge. 
Based on this formalism, we have implemented a transport code which includes all reduced density matrix coherences between energetically degenerate state. The latter are necessary to capture the interference effects which characterize our system.

%%%%%%%%%%%%%%%%%%%%%%%%%%%%%%%%%%%%%%%%%%
\begin{figure*}[ht!]
    \includegraphics[width=500pt,draft=false]{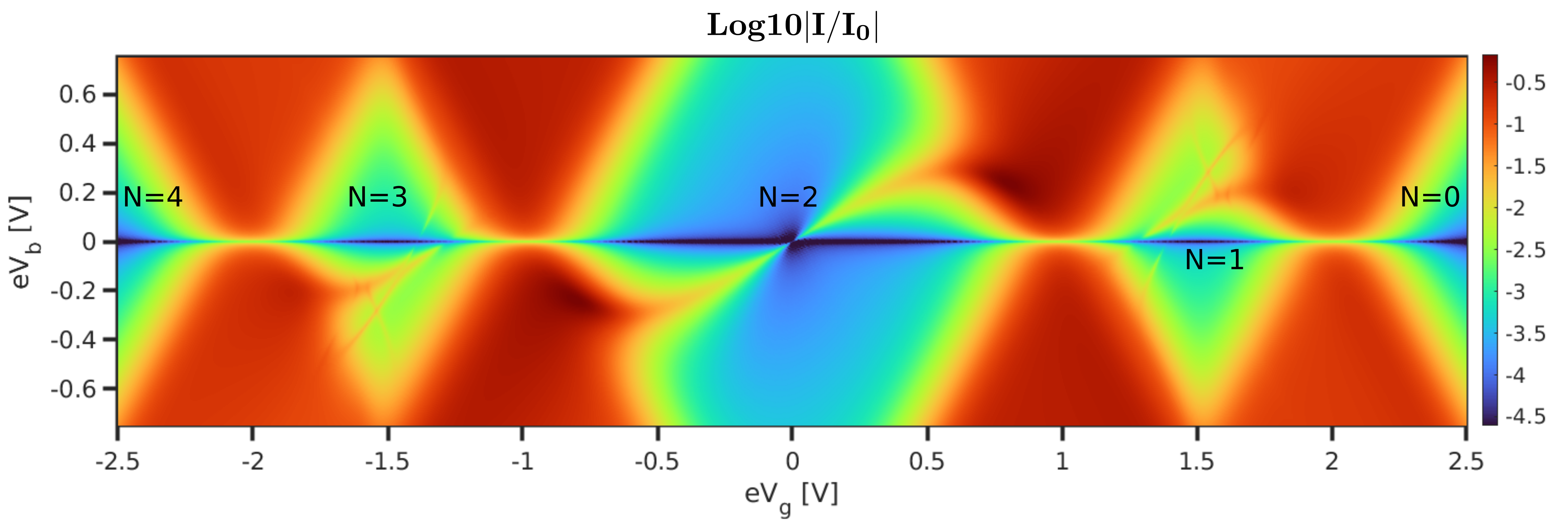} 
    \caption{\label{fig:overview_CD} Current plot of a DQD in a $V_\text{g}$-$V_\text{b}$ map shows an intricate set of current resonances: The $N=1,2,3$-Coulomb diamonds are decorated with resonances which cut deep into the Coulomb blockade regions. In the central $N=2$-diamond a simple ground-state-to-ground-state-transition is appearing. The parameters are the following: $U=2\,V$, $k_\text{B}T=0.05\,V$, $P_T=0.6$, $P_S=0.99$, $\theta=1.5$, $\phi=0.95 \pi$, $\Gamma^\text{L}_0=1 \times 10^{-2}\,V=2 \Gamma^\text{R}_0$ and $\varepsilon_0=-2\,V$. }
\end{figure*}
%%%%%%%%%%%%%%%%%%%%%%%%%%%%%%%%%%%%%%%%%%

\section{Numerical results}
\label{sec:numerical_results}
In Fig.~\ref{fig:overview_CD}, we show the current through the DQD calculated, according to the transport theory presented in the previous section, within the cotunneling limit. In particular, we set parallel pseudospin polarization with $P_T=0.6$ and pseudospin polarization angle with respect to the $z$ hard axis of $\theta=1.5$, as well as $P_S=0.99$ with a relative spin polarization angle $\phi=0.95\pi$. 

The current, given in logarithmic scale, is normalized to a reference value $I_0$. The latter is the one expected for a quantum dot spin valve in the high bias limit, but without pseudospin polarization \cite{Braun2004}. In terms of the system parameters introduced in the previous section, we calculate  $I_0=2(\Gamma_0^\text{L}\Gamma_0^\text{R})/(\Gamma_0^\text{L}+\Gamma_0^\text{R})\{1-[P_S \sin (\phi/2)]^2\}$. Such a current normalization highlights the effects of the pseudospin on the transport characteristics, since $I_0$ gives the scale of the underlying spin valve suppression.

The stability diagram is characterized, on the large scale, by five Coulomb diamonds where the Coulomb interaction suppresses the current, thus stabilizing a constant charge on the system. The quantized occupation of the DQD increases from 0 to 4 electrons by lowering the single-particle level, as indicated on the figure. The size $U$ and $V$ for, respectively, the two- and one- or three-particle Coulomb diamonds is determined by the corresponding addition energies. Besides of the electron-electron interaction, the current in the Coulomb diamonds is further suppressed, at biases larger than the temperature, by the spin valve configuration, which promotes spin accumulation on the system with an orientation antiparallel to the one of the drain lead.  

A distinctive current resonance protrudes into the Coulomb blockade area of the central diamond. It is a spin resonance which lifts the additional current suppression due to the spin valve configuration. We rationalize such a resonance, in the same spirit of \cite{Hell2015,Rohrmeier2021}, by introducing the exchange field:
\begin{equation}
    \label{eq:exchange_field_S2}
        \vec{B}_{2S}= \sum_l 2 P_S \Gamma^l_0 \left[p_l\!\left(E_{32g}\right)-p_l\!\left(E_\text{2g1}\right)\right] \vec{n}^l_S
\end{equation}
with 
\begin{equation}
    p_l\!\left(x\right)=\frac{\text{Re} \Psi^{(0)} \big( \frac{1}{2}+i\frac{ x- \mu_l}{2 \pi k_\text{B}T}\big)}{2\pi},
\end{equation}
where $\Psi^{(0)}(z)$ is the digamma-function, $T$ the temperature and $k_\text{B}$ the Boltzmann factor. The subscript of the energy $E_{\text{s}\text{s}'}$ labels the energy difference between the many-body eigenstates $\text{s}$ and $\text{s}'$. 

The electronic fluctuations from and to the leads, in combination with the Coulomb interaction on the system, are at the origin of this exchange field. The latter is only effective in presence of a degenerate energy spectrum. Analogously to an external magnetic field, it generates a spin procession on the triplet sector of the two-particle ground state (cf.\ Fig.~\ref{fig:setup2}). This mechanism counteracts the spin accumulation and lifts the spin valve suppression by promoting precession towards the spin states more aligned to the drain polarization. The position of the resonance can thus be predicted by the vector condition $\vec{B}_{2S} \cdot \left(\vec{n}^\text{L}_S-\vec{n}^\text{R}_S\right)=0$ \cite{Hell2015,Maurer2020,Rohrmeier2021}. The latter defines, in fact, the exchange field which maximizes the precession towards the drain spin direction.

Also the one- and three-particle diamonds are decorated by current resonances. Their pattern is, though, more intricate than the one of the central Coulomb diamond and it cannot be explained solely in terms of exchange field induced spin precession. The explanation requires a more detailed analysis involving the interplay with the pseudospin degree of freedom.  Due to the particle-hole symmetry of the Hamiltonian, we will restrict ourselves to the one-particle diamond. Results for negative energies can be deduced by a simultaneous reflection of both the bias and the gate voltage.     

\begin{figure*}
    \includegraphics[width=500pt,draft=false]{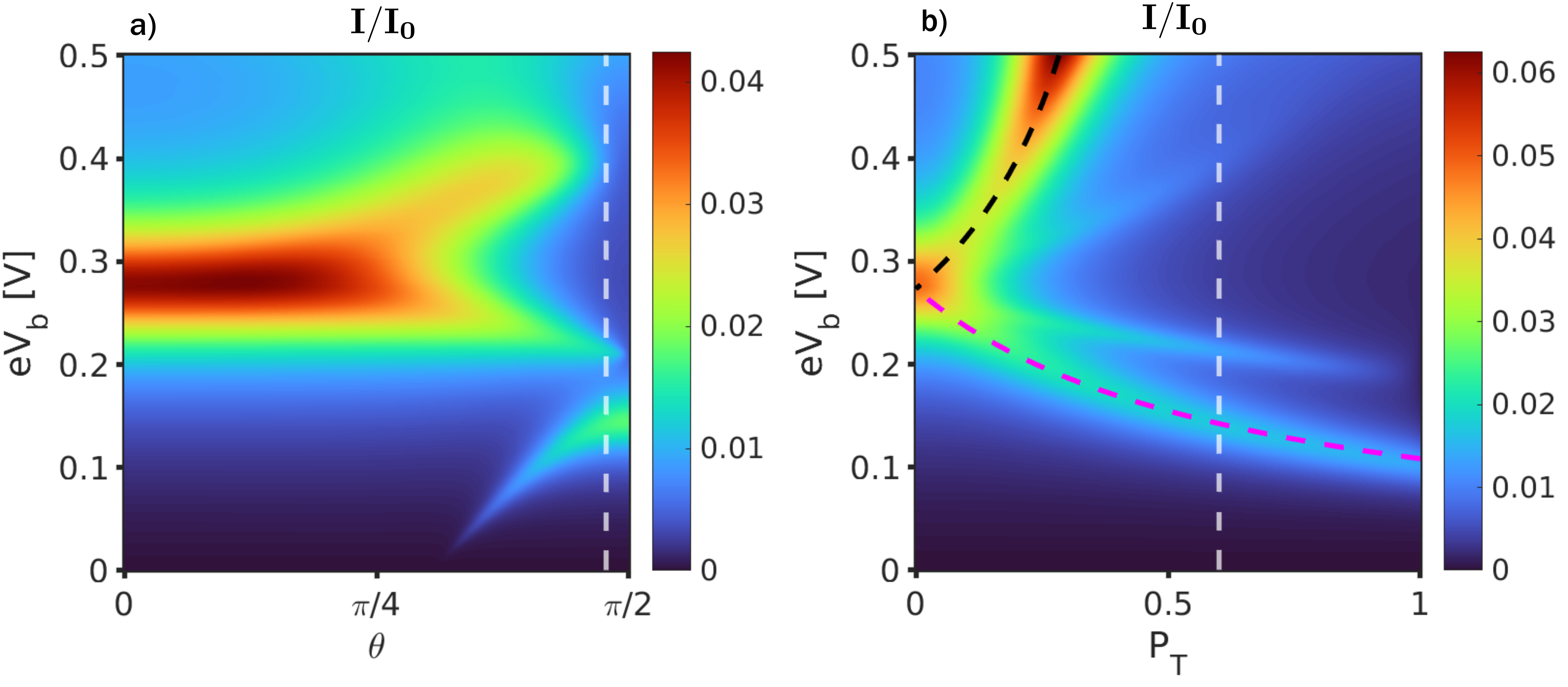} 
    \caption{\label{fig:pol_angle_plot} Current resonances modulated by the pseudospin polarization angle $\theta$ and strength $P_T$: a) $\theta$-$V_\text{b}$ map and b) $P_T$-$V_\text{b}$ map, both at $V_\text{g}=1.5\,V$, exhibit a strong dependence on the respective parameters. The white dashed lines indicate the parameter set of Fig.~\ref{fig:overview_CD}. The black and magenta dashed lines are the resonance conditions for the $\vec{S}_-$ and $\vec{S}_+$ channel, which are explicitly discussed in Sec.~\ref{sec:limiting_cases}.}
\end{figure*}

The intimate relation between the current resonances of the one-particle diamond and the pseudospin degree of freedom is presented in Fig.~\ref{fig:pol_angle_plot}. The current resonances are plotted here as a function of the pseudospin polarization $P_T$ and the pseudospin polarization angle $\theta$. Not only the position and the strength of the resonances, but even their number, depends on the control parameters. For example, the angle dependence shows a single peak for $\theta=0$ which splits into two and even acquire a third resonance for larger angles. The mere $z$ polarization of the leads for $\theta = 0$, allows, to identify parallel transport channels for each of the dots and could be rationalized by a spin exchange field similar to Eq.~\eqref{eq:exchange_field_S2}. The same procedure, though, fails to capture all the resonances for intermediate angles $0<\theta<\pi/2$ and intermediate polarization strengths $0<P_T<1$. 

For a more complete understanding of the entire parameter range, we introduce, in the next section, a reduced model and study the dynamics of the system within the lowest order in the tunneling coupling.

%%%%%%%%%%%%%%%%%%%%%%%%%%%%%%%%%%%%%%%%%%
%%%%%%%%%%%%%%%%%%%%%%%%%%%%%%%%%%%%%%%%%%

\section{Coherent sequential tunneling model}
\label{sec:CST_model}
We deduce the equations for the minimal model by considering only the sequential tunneling contributions to the full GME. For simplicity, we perform also the Markov approximation which, anyway, does not influence the stationary solution. Under these conditions, the equation of motion for the reduced density operator reads:
\begin{equation}
\label{eq:GME_2order}
    \dot{\hat{\rho}}_{\rm red} = -\frac{i}{\hbar}[\hat{H}_{\rm DQD} + \hat{H}_{\rm LS},\hat{\rho}_{\rm red}] + \lio_{\rm T}\hat{\rho}_{\rm red},
\end{equation}
where $\lio_{\rm T}$ describes the tunneling events among many-body states with consecutive particle numbers and $ \hat{H}_{\rm LS}$ is the Lamb shift Hamiltonian, which renormalizes the coherent DQD dynamics and is due to virtual charge fluctuations \cite{Donarini2010}. A detailed derivation of Eq.~\eqref{eq:GME_2order} for the model at hand is given, for example, in \cite{Rohrmeier2021}, where, though, a completely different parameter regime has been analyzed.

The richest pattern of anomalous current resonances is found in the one-particle Coulomb diamond (cf. Fig.~\ref{fig:overview_CD}). Thus, we further restrict ourselves only to the elements of the density matrix describing the empty and the single-occupied DQD.

The system exhibits a fourfold degenerate one-particle spectrum and single-particle tunneling rate matrices which cannot be diagonalized simultaneously. Thus, in general, all the one-particle coherences should be retained for a correct description of interference effects \cite{Donarini2010}. Their dynamics is, in fact, coupled to the one of the corresponding populations, independently of the representation basis. Such a transport regime goes under the name of coherent sequential tunneling \cite{Maurer2020}.      

In summary, the non-equilibrium dynamics of the DQD weakly coupled to the source and drain leads reduces, in the above-mentioned limit, to a set of 17 coupled linear differential equations, involving the empty-state population and each of the 16 elements of the one-particle density matrix. 

Alternatively, a rather standard description involves the expectation values of a complete set of operators: $P_0 = \langle \hat{P}_0\rangle$, $P_1 = \langle \hat{P}_1\rangle$, $T_{\alpha}= \langle \hat{T}_\alpha\rangle$, $ S_{\alpha} = \langle \hat{S}_\alpha\rangle$ and $\Lambda_{\alpha \beta} = \langle \hat{T}_\alpha \hat{S}_\beta \rangle$ where $\hat{P}_0 = |\emptyset\rangle\!\langle\emptyset|$ is the projector on the empty state, $\hat{P}_1 = \sum_{i\sigma}\hat{d}^\dagger_{i\sigma}\hat{P}_0 \hat{d}_{i\sigma}$, $\hat{T}_\alpha$ is the $\alpha = x,y,z$ component of the pseudospin operator defined in Eq.~\eqref{eq:Pseudospin}, and analogously $\hat{S}_\alpha = 1/2\sum_{i \tau \tau'}  \hat{d}^\dagger_{i\tau} \sigma^\alpha_{\tau \tau'}  \hat{d}_{i\tau'}$. Within these 17 linearly independent variables, we replace, in this manuscript, $\vec{S} $ and $\Lambda$ by the four vectors:
\begin{equation}
    \label{eq:observables}
       \vec{S}_{\pm}  = \frac{\vec{S}}{2}\pm \et \cdot  \Lambda,  \quad 
       \vec{\Lambda}_\perp  =\eperp \cdot \Lambda, \quad
       \vec{\Lambda}_y  =\ey \cdot \Lambda, 
\end{equation}
which involve the orthogonal basis $\ey = (0,1,0)$, $\et = \vec{n}_T$ and $\eperp=(\ey \times \et)$. This basis adapts to the orientation of the (parallel) pseudospin polarization of the leads and analogously it occurs to the set of variables in Eq. \eqref{eq:observables} chosen to describe the system. The equation of motion for such observables read:
\begin{widetext}
    \begin{equation}
        \begin{split}\label{eq:eqofmodel}
            %%%%%%               P_0            %%%%%%
            &\dot{P_0} =  + \gamma^- \left( P_1 +2 P_T \et \cdot \tvec \right) - 4 \gamma^+ P_0  
            + 2 D_+ \vec{\gamma}^- \cdot  \vec{S}_+ +2 D_- \vec{\gamma}^- \cdot  \vec{S}_- ,\\
            %%%%%%               P_1            %%%%%%
            &\dot{P_1} = - \gamma^- \left(P_1 + 2 P_T \et \cdot  \tvec \right) + 4 \gamma^+ P_0   
            - 2 D_+ \vec{\gamma}^- \cdot  \vec{S}_+ -2 D_- \vec{\gamma}^- \cdot  \vec{S}_- = -\dot{P_0},\\
            %%%%%%               T            %%%%%%
            &\dot{\vec{T}}  = -\gamma^-\tvec +2 \vec{\omega}_{T} \times \tvec 
            -\left[\gamma^-\tfrac{P_T}{2} P_1-2\gamma^+ P_T P_0  +  D_+  \vec{\gamma}^- \cdot  \vec{S}_+  - D_-  \vec{\gamma}^- \cdot \vec{S}_-  -4 \vec{\omega}_{S}^{\rm a} \cdot \vec{\Lambda}_y  \right]\et    \\
            & \quad \quad +\left[4 \vec{\omega}_- \cdot  \vec{\Lambda}_\perp  -2  \vec{\omega}_{S}^{\rm a}  \cdot (\vec{S}_+-\vec{S}_-) -2 \vec{\gamma}^- \cdot \vec{\Lambda}_y  \right]\ey -\left[ 4 \vec{\omega}_- \cdot \vec{\Lambda}_y   +2 \vec{\gamma}^- \cdot \vec{\Lambda}_\perp   \right]\eperp  , \\
             %%%%%%               Splus/minus          %%%%%%
            &\dot{\vec{S}}_\pm =  - \gamma^- D_\pm \vec{S}_\pm +2\left(\vec{\omega}_S \pm \vec{\omega}_- \right) \times  \vec{S}_\pm + D_\pm\left[ \vec{\gamma}^+ P_0 - \tfrac{\vec{\gamma}^-}{4} (P_1\pm 2\et \cdot\tvec)   \right]   +2 \vec{\omega}_{S}^{\rm a} \times \vec{\Lambda}_\perp \mp2 \omega_T^{\rm a} \vec{\Lambda}_y \pm \vec{\omega}_{S}^{\rm a} (\ey \cdot \tvec)  ,   \\
            %%%%%%               Splus           %%%%%%
           % \langle \dot{\vec{S}}_+ \rangle =& - \left(1+P_T\right) \left[ \gamma^- \splus + \vec{\gamma}^-\left(\tfrac{P_1}{4}+ \et \cdot \tfrac{\tvec}{2} \right) - \vec{\gamma}^+ P_0 \right] \nonumber \\
           % &+2\left(\vec{\omega}_S+\vec{\omega}_- \right) \times \splus +2 \sin \theta \left(\vec{\omega}_{TS} \times \tperps - \omega_T^{an} \tys + \ey \cdot  \tfrac{\tvec}{2} \vec{\omega}_{TS}   \right) \\
            %%%%%%               Sminus           %%%%%%
           % \langle \dot{\vec{S}}_- \rangle =& - \left(1-P_T\right) \left[ \gamma^- \sminus + \vec{\gamma}^-\left(\tfrac{P_1}{4}- \et \cdot \tfrac{\tvec}{2} \right) - \vec{\gamma}^+ P_0 \right] \nonumber \\
           % &+2\left(\vec{\omega}_S - \vec{\omega}_- \right) \times \sminus +2 \sin \theta \left(\vec{\omega}_{TS} \times \tperps + \omega_T^{an} \tys- \ey \cdot  \tfrac{\tvec}{2} \vec{\omega}_{TS}  \right) \\
             %%%%%%               TperpS           %%%%%%
            &\dot{\vec{\Lambda}}_\perp  = -\gamma^-\vec{\Lambda}_\perp + 2\vec{\omega}_S \times \vec{\Lambda}_\perp 
            -2 \omega_+ \vec{\Lambda}_y - P_T \vec{\gamma}^- \times \vec{\Lambda}_y - \vec{\omega}_- (\ey \cdot \tvec)   
            - \tfrac{\vec{\gamma}^-}{2} \eperp \cdot \tvec +  \vec{\omega}_{S}^{\rm a} \times (\vec{S}_+ +\vec{S}_-),  \\
              %%%%%%               TyS           %%%%%%
            &\dot{\vec{\Lambda}}_y =  -\gamma^-\vec{\Lambda}_y  +2\vec{\omega}_S \times \vec{\Lambda}_y
            +2 \omega_+ \vec{\Lambda}_\perp + P_T \vec{\gamma}^- \times \vec{\Lambda}_\perp +\vec{\omega}_- (\eperp \cdot \tvec)  
            - \tfrac{\vec{\gamma}^-}{2} \ey \cdot \tvec   - \vec{\omega}_{S}^{\rm a} (\et \cdot \tvec)         +  \omega_{T}^{\rm a} ( \vec{S}_+-\vec{S}_-),
        \end{split}
\end{equation}
\end{widetext}
where several functions have been defined to express the tunneling, as well as the Lamb shift contribution of the Liouvillian.  On one hand we have introduced scalar and vector rates, respectively
\begin{equation}
\gamma^\pm=\sum_l \gamma^\pm_l \quad \text{with} \quad \gamma^\pm_l = \frac{\Gamma^l_0}{4} f_l^\pm(\varepsilon)   , 
\end{equation}
and 
\begin{equation}
    \vec{\gamma}^\pm= \sum_l \vec{\gamma}^\pm_l \quad  \text{with} \quad 
    \vec{\gamma}^\pm_l = P_{S} \vec{n}^l_S \gamma^\pm_l ,
\end{equation}
in which, for the Fermi functions, we adopt the notation $f_l^\pm (\varepsilon) = [e^{\pm(\varepsilon-\mu_l)/(k_\text{B}T)}+1]^{-1}$. Furthermore, we set $D_\pm = 1 \pm P_T$ to quantify the coupling strength to the different pseudospin sectors. 

The Lamb shift contribution to the GME yields several exchange fields, which are responsible for precession dynamics for the vectorial components in Eq.~\eqref{eq:eqofmodel}. To this end, we introduce the frequencies $\omega_{xx',yy'}^l= \Gamma^l_0 [ p_l(E_{xx'})-p_l(E_{yy'}) ]/4$, which involve the difference of two digamma-functions, and fluctuations towards both the zero- and the two-particle neighboring states. 
In terms of those frequencies, we define the exchange fields:
\begin{equation}
\label{eq:exchange_fields}
    \begin{split}
        \vec{\omega}_T&=P_T\sum_l (\omega^l_{10,2g1} \vec{n}_T+ \omega^l_{2e1,2g1} \cos{\theta}\, \ez), \\
        \vec{\omega}_S&=P_S\sum_l \omega^l_{10,2e1} \vec{n}_S^l ,\\
        \vec{\omega}_-&=P_T P_S\sum_l  (\omega^l_{10,2g1} - \omega^l_{2e1,2g1} \cos^2 \theta) \vec{n}_S^l, \\
        \vec{\omega}_S^\text{a}&=P_T P_S \sin{\theta} \cos{\theta}\sum_l  \omega^l_{2e1,2g1} \vec{n}_S^l. 
    \end{split}
\end{equation}
The list of auxiliary functions appearing in the model equations \eqref{eq:eqofmodel} is complemented by the scalars
\begin{equation}
\label{eq:exchange_scalars}
    \begin{split}
        \omega_T^\text{a}&=P_T \sin{\theta} \cos{\theta} \sum_l   \omega^l_{2e1,2g1}, \\
        \omega_+&=P_T \sum_l  (\omega^l_{10,2g1} + \omega^l_{2e1,2g1} \cos^2 \theta).
    \end{split}
\end{equation}
Despite their complexity, Eqs.~\eqref{eq:eqofmodel} display simple recurring patterns, which can guide us in the understanding of their physical implications. 

The first two equations express the rate of change in the zero- and one-particle populations. It holds, in particular, $\dot{P}_0 = -\dot{P}_1$ as follows from the probability conservation and the neglect of populations with particle number larger than one. This assumption strongly reduces the number of equations needed to describe the DQD, but it also restricts their validity to the region of the one-particle diamond closer to the $0-1$ charge degeneracy point. 

The rate of change of $P_0$ (and $P_1$) not only depends on populations, but also on the spin and pseudospin vectors, respectively  $\vec{S}_\pm$ and $\vec{T}$. The latter appear within scalar products with, respectively, the spin and the pseudospin polarization vectors in the leads. Indeed, like for a spin valve, electron tunneling is favoured by the alignment of the spin or the pseudospin degree of freedom of the DQD with the corresponding lead polarization.

We now turn to the equation of motion for the vectorial components in Eqs. \eqref{eq:eqofmodel}, which  all share the same structure and encompass three main effects: decoherence, precession and pumping. The first two effects are described by the terms involving the very same vector whose time derivative appears on the LHS of the equation. We collect instead under the concept of pumping all the other  terms, involving populations as well as the other vectors describing the DQD. 

The rate of decoherence is always proportional to $\gamma^-$ as tunneling events towards the zero-particle state reduce both the spin as well as the pseudospin on the DQD. Such processes, though, are strongly suppressed within the one-particle diamond, due to Coulomb interaction. We thus expect weak decoherence. Even the correlator vectors $\vec{\Lambda}_y$ and $\vec{\Lambda}_\perp$ are subject to the same decoherence rate. Notice, moreover, the $D_\pm$ weight is modulating the rate of the spin variables $\vec{S}_\pm$, which implies a further reduction of decoherence for the spin variable $\vec{S}_-$ in presence of large pseudospin polarization. 

The exchange fields characterizing the precession terms strongly vary, among the different vectorial components, both in direction and intensity. The pseudospin exchange field always points into the direction $\et$, i.e. the one of the parallel pseudospin polarizations of the leads.  The spin exchange field results instead from a delicate balance between the almost antiparallel source and drain contributions. Thus, both the strength and the intensity of the fields $\vec{\omega}_S \pm \vec{\omega}_-$ are strongly modulated within the one-particle Coulomb diamond.   

The pumping component of the (pseudo)spin dynamics is the one responsible for the (pseudo)spin accumulation on the DQD observed in the stationary limit. Naturally, such a phenomenon characterizes the spin channels, due to the spin valve configuration of the leads polarization. The spin pumped from the source lead accumulates, in absence of spin precession, on the DQD and it has hardly any chance to escape towards the almost antiparallel polarized drain. The terms encompassing this dynamics are the ones proportional to the populations $P_0$ and $P_1$. The pumping component contains, moreover, also terms which intertwine the spin dynamics to the one of the  pseudospin and that of the correlator vectors $\vec{\Lambda}_y$ and $\vec{\Lambda}_\perp$. Analogously, thanks to the coupling to the other vectorial variables, also the pseudospin can be pumped along a generic direction, despite the parallel polarization of the leads along $\et$. 

The effects of such an intricate system dynamics on the transport characteristics and, in particular, the crucial role played by the spin and the pseudospin degree of freedom is illustrated by the current formula:
\begin{equation}
    \begin{split}\label{eq:current_model}
        I_{\rm model} =& 4\left(\gamma^+_L-b\gamma^+\right) -2P_T\left( \gamma^-_L - b \gamma^-\right) \vec{n}_T \cdot \vec{T}^\infty  \\
        & -2 \left( \vec{\gamma}^-_L - b \gm \right) \cdot \left(D_+ \vec{S}_+^\infty + D_- \vec{S}_-^\infty\right),
    \end{split}
\end{equation}
in which $b= (\gamma^-_L+4\gamma^+_L)/(\gamma^-+4\gamma^+)$ and the superscript "$\infty$" indicates observables calculated in the steady state limit. 

\begin{figure}
    \includegraphics[width=\columnwidth,draft=false]{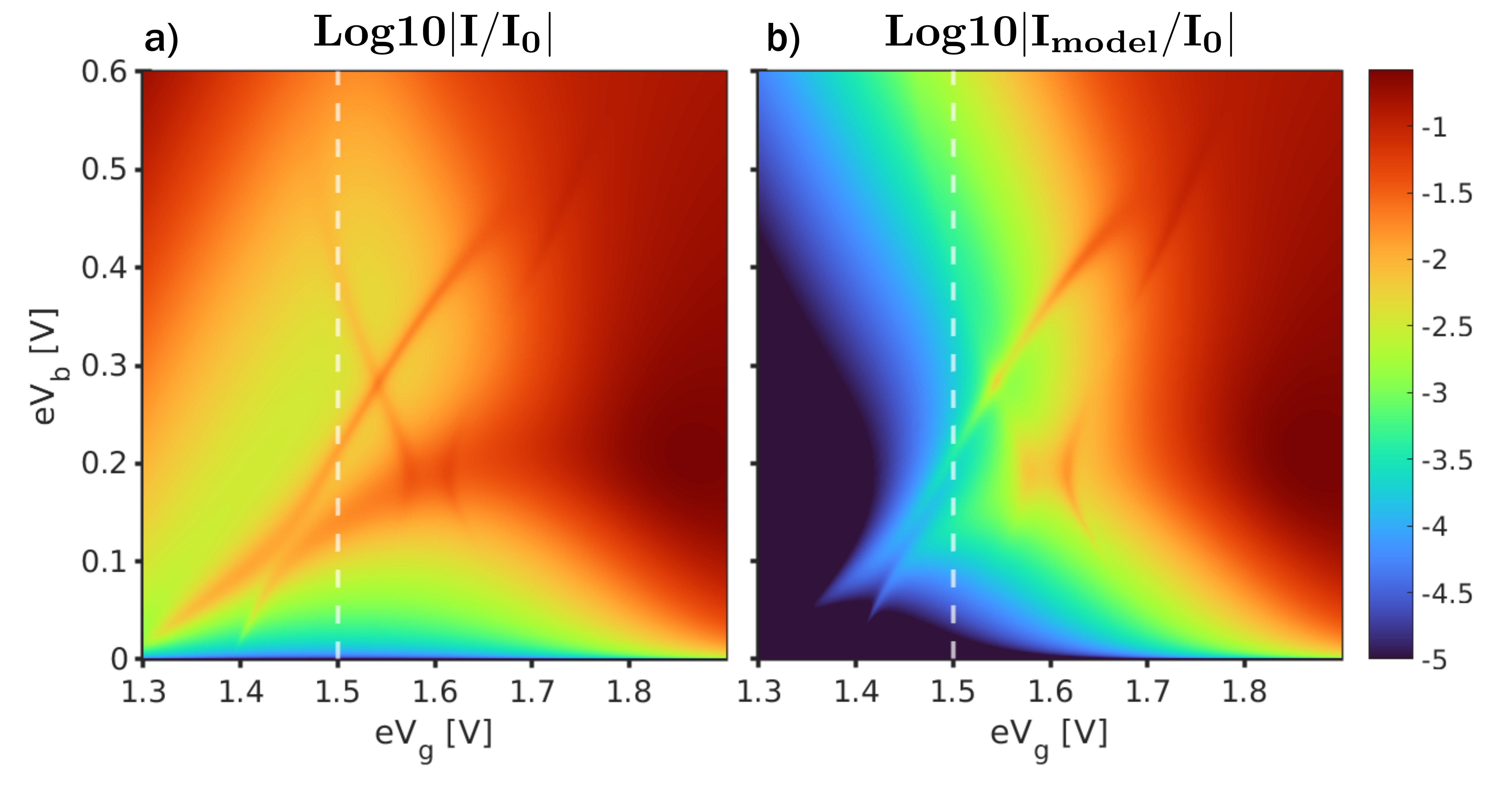}
    \caption{\label{fig:current_comparison} Current in the one-particle Coulomb diamond calculated with two different approaches: In panel a) the full cotunneling calculation is presented. Panel b) shows the corresponding result in the  coherent sequential tunneling limit. Both currents are renormalized by the current $I_0$ expected for a spin valve in the high bias limit. The white dashed line should help for the comparison with Fig.~\ref{fig:pol_angle_plot}.}
\end{figure}

A comparison between the current in the one-particle Coulomb-diamond obtained in the cotunneling approximation with the one stemming from this coherent sequential tunneling model is depicted in Fig.~\ref{fig:current_comparison}. Despite the strong simplifications in the model calculation, the two currents show a good qualitative agreement. In particular, the main resonance which is bending towards the point ($V_\text{g}\approx1.3$,$V_\text{b}=0$) as well as its anti-crossing near the point ($V_\text{g}\approx1.4, V_\text{b}\approx0.1$) are captured  in the model description. Of the distinctive cross-shaped feature of the cotunneling calculation, though, only one arm is well visible in the model calculation. The other arm is buried inside the fermionic tail of the current inside a Coulomb diamond and therefore it is  barely discernible. Moreover, a poorer match is expected for the side of the Coulomb diamond closer to the 1-2 charge degeneracy point. As we neglect, for simplicity, direct tunneling to the two-particle states, the current of the model decreases exponentially for decreasing gate voltages.

The first component in Eq.~\eqref{eq:current_model} yields the current expected for $P_S = P_T = 0$. As it only contains Fermi functions centered around the $0-1$ transition resonance, this contribution to the current is smooth within the one-particle Coulomb diamond. Consequently, the sharp current resonances observed in Fig.~\ref{fig:current_comparison}b) can only be ascribed to the sharp modulations of the stationary pseudospin and spin vectors appearing respectively in the second and third term of Eq.~\eqref{eq:current_model}. 

\begin{figure*}
    \includegraphics[width=450pt,draft=false]{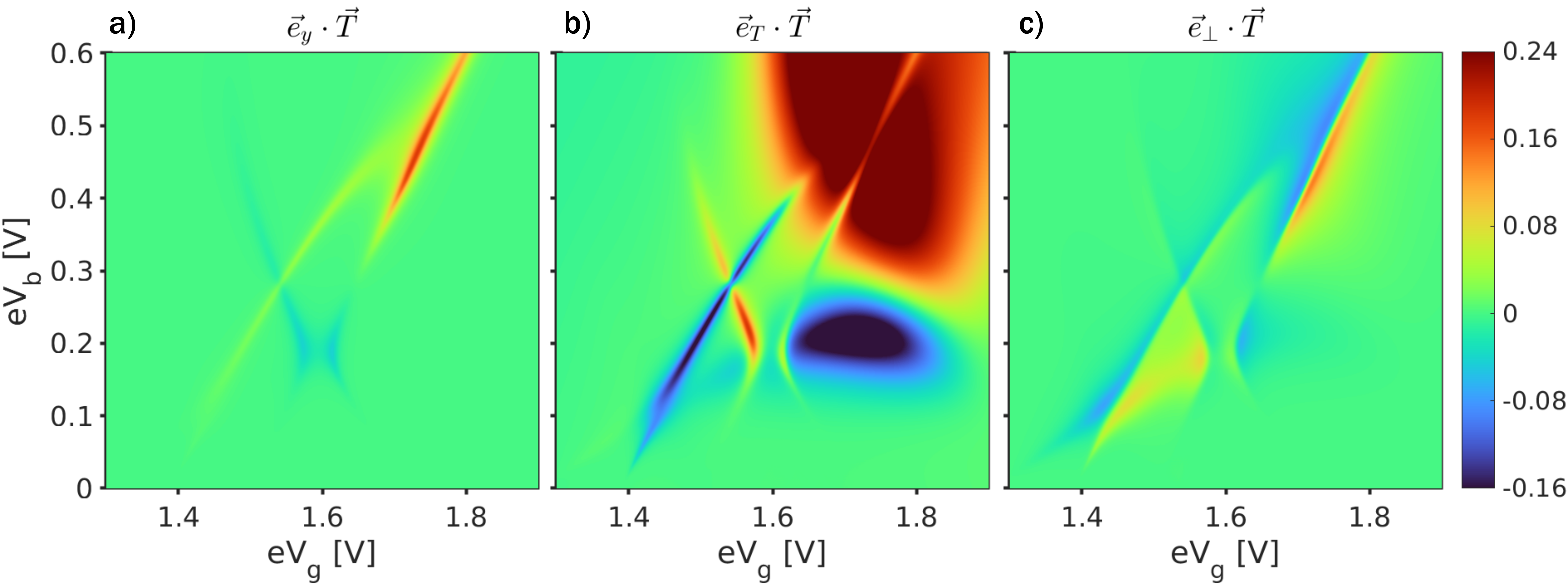} 
    \caption{Pseudospin depends strongly on the gate and bias voltages: The components a) $\ey$, b) $\et$, c) $\eperp$ of $\vec{T}$ underline the vector character of the pseudospin. The (anti-)alignment of the pseudospin (decreases) increases the current flow through the DQD. Focusing on the upper right corner, one observes a clear rotation of $\vec{T}$ in the $y$ direction. Same parameters as in Fig.~\ref{fig:overview_CD}.   \label{fig:pseudospin} }
\end{figure*}

In Fig.~\ref{fig:pseudospin}, the components of $\vec{T}$ in the basis of  $\ey$, $\et$, and $\eperp$ are displayed. Distinct features in the pseudospin components are clearly correlated to the current resonances in Fig.~\ref{fig:current_comparison}. For most of the bias and gate voltages, $\vec{T}^\infty$ points along the $\et$ direction and in the areas of (anti-)alignment of $\vec{T}$ with respect to $\et$, the current is (lowered) raised. There are, though also areas in which the other components of $\vec{T}$ prevail and the pseudospin contribution to the current vanishes, as can be derived from Eq.~\eqref{eq:current_model}. Altogether, it is thus clear how the vectorial character of $\vec{T}$ must be considered for a thorough description of the transport phenomena. In particular, it is the intertwining of the spin and pseudospin degrees of freedom which foster the drastic deviation of the pseudospin direction of the DQD from the polarization direction of the leads, as will be highlighted later by analyzing limiting cases of the pseudospin polarization angle $\theta$.

%%%%%%%%%%%%%%%%%%%%%%%%%%%%%%%%%%%%%%%%%%
%%%%%%%%%%%%%%%%%%%%%%%%%%%%%%%%%%%%%%%%%%

%%%%%%%%%%%%%%%%%%%%%%%%%%%%%%%%%%%%%%%%%%
%%%%%%%%%%%%%%%%%%%%%%%%%%%%%%%%%%%%%%%%%%

We now turn to the spin contribution of the current. The first qualitative understanding is obtained in the framework of the phenomenology of a quantum dot spin valve. The last term of Eq.~\eqref{eq:current_model} substantially decreases the current due to the almost antiparallel alignment of the source and drain and the corresponding spin accumulation along the source spin polarization direction.

More specifically, we refer in Eq.~\eqref{eq:current_model} to the combinations of the spin vector $\vec{S}$ and the spin-pseudospin correlator $\Lambda$ proposed in Eq.~\eqref{eq:observables}. The latter define spin observables which, for specific limiting cases, identify independent spin channels. The full separation is only obtained when $\vec{n}_T$ coincides with the hard axis ($\theta = 0$) or it belongs to the easy plane ($\theta = \pi/2$) for the pseudospin of the DQD and will be discussed in detail in Sec.~\ref{sec:limiting_cases}. Insight into the spin dynamics can though be gained also for the case at hand ($\theta = 1.5$) as it is demonstrated by the predicting character of the magenta and black dashes lines in Fig.~\ref{fig:pol_angle_plot}b).

%%%%%%%%%%%%%%%%%%%%%%%%%%%%%%%%%%%%%%%%%%
\begin{figure}
    \includegraphics[width=\columnwidth,draft=false]{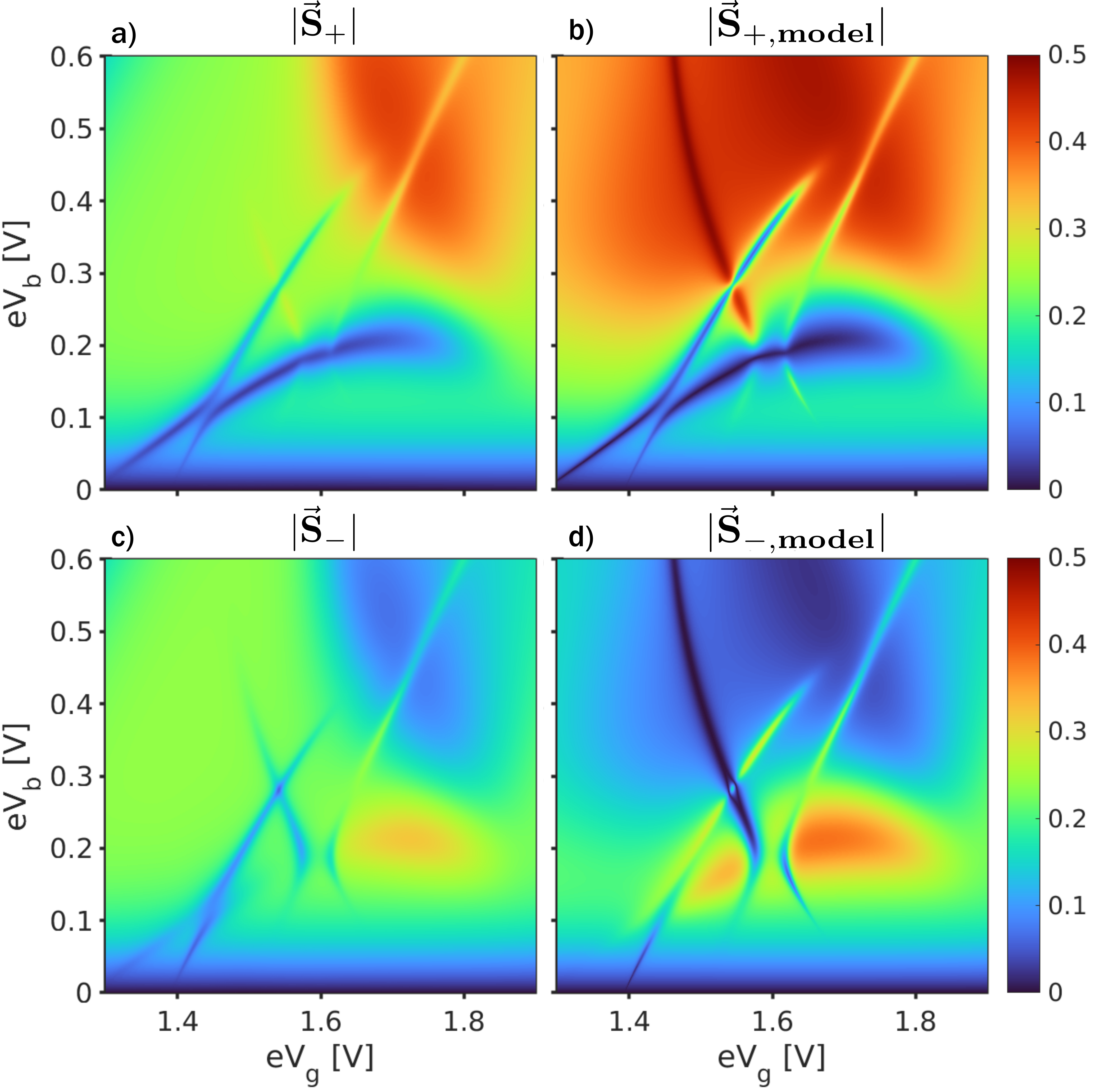} 
    \caption{Comparison of stationary spin variables $\vec{S}_\pm$: Panel a) and c) are obtained from the full cotunneling calculation, while b) and d) refer to the reduced sequential tunneling model. Same parameters as in Fig.~\ref{fig:overview_CD}.
\label{fig:spin_dynamics} }
\end{figure}
%%%%%%%%%%%%%%%%%%%%%%%%%%%%%%%%%%%%%%%%%%

In Fig.~\ref{fig:spin_dynamics}, we compare the modulus of the two spin variables $\vec{S}_+$ and $\vec{S}_-$ as calculated from the full cotunneling and from the coherent sequential tunneling model. The model captures the rich texture of the stationary spins even better as compared to the current  presented in Fig.~\ref{fig:current_comparison}. 
Particularly, the cross-shaped feature appears more distinctively, although for $|\vec{S}_-|$ the model predicts the wrong sign for the positive slope signal. Most interestingly, we observe how the spin channels can be blocked or unblocked individually, as the separate regions of high, respectively, low modulus indicate. Qualitatively, we can rationalize this phenomenon as a transfer of probability between the "+" and the "-" channel occurring when one of them is unblocked due to a fast precession dynamics. 

A more quantitative description is obtained, analyzing the equation of motion for $\vec{S}_\pm$. The latter can be divided into a decoherence, a precession and a pumping component:
\begin{equation}
\dot{\vec{S}}_\pm  =  
\underbrace{-a_\pm \vec{S}_\pm}_{\rm decoherence} 
+
\underbrace{\vec{B}_\pm \times  \vec{S}_\pm}_{\rm precession}
+ 
\underbrace{\vec{x}_\pm(P_1, \tvec, \vec{\Lambda}_y, \vec{\Lambda}_\perp)}_{\rm pumping}.
\end{equation}
The steady state solution of this equation is given by $\vec{S}^\infty_\pm= \vec{F}(a_\pm, \vec{x}_\pm, \vec{B}_\pm)$ with
\begin{equation}\label{eq:ffunction}
    \vec{F}(a,\vec{x},\vec{B}) = \frac{a}{a^2+|\vec{B}|^2}\left(\vec{x}+\frac{\vec{B} \cdot \vec{x}}{a^2}\vec{B}+\frac{\vec{B} \times \vec{x}}{a}\right).
\end{equation}
We define the input parameters as $a_\pm=D_\pm \gamma^-$ and $\vec{B}_\pm = 2 \left( \vec{\omega}_S \pm \vec{\omega}_- \right)$. Furthermore, we use for the pumping the steady state solution of the other variables:
\begin{equation}
\label{eq:pumping_S}
    \begin{split}
    \vec{x}_\pm&= D_\pm \left[\vec{\gamma}^+(1-P_1^\infty)  - \frac{\vec{\gamma}^-}{4}(P_1^\infty \pm 2\et \cdot\tvec^\infty) \right]\\
    &+2 \vec{\omega}_{S}^{\rm a} \times \vec{\Lambda}_\perp^\infty 
    \mp 2\omega_T^{\rm a} \vec{\Lambda}_y^\infty \pm  \vec{\omega}_{S}^{\rm a} (\ey \cdot \tvec^\infty)   .      
    \end{split}
\end{equation}
We do not have a closed form solution of the intricate Eqs.~\eqref{eq:eqofmodel}. The analysis of the semi-analyitical Eq.~\eqref{eq:ffunction} gives, though, relevant insights on the accumulation dynamics of the spin variables.

We distinguish among three different regimes, depending on the ratio of $|\vec{B}_\pm|/a_\pm$. If the decoherence rate is much larger than the precession frequency ($|\vec{B}_\pm|/a_\pm\ll1 $) the respective stationary spin is given by $\vec{S}_\pm\approx \vec{x}_\pm/a_\pm$, at most, corrected by the small precession contribution $\vec{B}_\pm \times \vec{x}_\pm/a_\pm^2$. Essentially, the pumping defines the accumulation direction. 

The opposite regime is obtained whenever ($|\vec{B}_\pm|/a_\pm \gg 1 $). In this case, the second term of Eq.~\eqref{eq:ffunction} dominates and results in dephasing, with all components suppressed except for the ones pointing in the direction of the exchange fields. 

In the intermediate regime ($|\vec{B}_\pm|/a_\pm \approx 1 $), also the last term, which represents a coherent precession of the pumped spin, plays an important role. Inside the one-particle Coulomb diamond, it holds $|\vec{B}_\pm|/a_\pm \gg 1 $ so that the spin is mainly determined by the absolute value of the pumping $|\vec{x}_\pm|$ and the angle $\angle(\vec{x}_\pm,\vec{B}_\pm)$ between pumping direction and exchange field. 

If the pumping occurs for a given spin variable in a direction perpendicular to the exchange field, the corresponding spin is strongly dephased and, for that channel, the spin blockade is strongly lifted. Since the same condition cannot occur simultaneously for both spin channels, the other one absorbs probability. This probability transfer corresponds to an increase of the pseudospin component along $\et$. The latter, in turns, is also precessing (see Fig.~\ref{fig:pseudospin}) and it gives feedback on the spin pumping direction. 

While the population transfer between the "+" and the "-" spin channels rationalizes the complementary behavior of the spin plots in Fig.~\ref{fig:spin_dynamics}, the interplay between the spin and the pseudospin is at the origin of the correlation between Figs.~\ref{fig:pseudospin} and \ref{fig:spin_dynamics}. 

All together, the two-spin-channel description represents a good starting point for unraveling the dynamics of the DQD spin valve under consideration. A fully vectorial approach to the pseudospin, going beyond the population difference of the spin channels (the latter being represented by the $\et\cdot\vec{T}$) is though necessary for a generic orientation of the pseudospin polarization.

\section{Limiting cases}
\label{sec:limiting_cases}

We consider, in this section, two limiting cases of pseudospin polarization direction: firstly, we assume with $\theta=0$ that $\et$ coincides with the hard pseudospin axis; afterwards, we take   $\et$ in the easy plane, i.e. $\theta=\pi/2$. The symmetry of the system Hamiltonian with respect to any rotation around the hard pseudospin axis ensures the equivalence of all pseudospin polarizations belonging to the easy plane. 

The fundamental simplification obtained for $\theta = 0$ or $\theta = \pi/2$ is the vanishing of the exchange field $\vec{\omega}_S^a$ as well as of the scalar $\omega_T^a$. Both functions derive from the Lamb shift contribution of the Liouvillian and, in particular, they originate from the pseudospin anisotropy of the DQD. Interestingly, for both limiting angles the variables $\vec{\Lambda}_\perp$, $\vec{\Lambda}_y$, are only coupled to themselves and to the  components $\ey \cdot \vec{T}$, $\eperp \cdot \vec{T}$ of the pseudospin, but they are independent of  $P_0$, $P_1$, $\vec{S}_+$, $\vec{S}_-$ and $\et\cdot\vec{T}$. If the system of Eqs.~\eqref{eq:eqofmodel} admits a unique stationary solution, the latter will correspond to the trivial choice for the set of coupled variables which do not include the populations. It is in fact the probability conservation to fix the normalization of the kernel for the Liouvillian. The relevant part of Eqs.~\eqref{eq:eqofmodel} can be cast with the help of $P_\pm=\tfrac{P_1}{2} \pm \vecnt \cdot \vec{T}$ into the following equations
\begin{equation}\label{eq:limiting_case}
        \begin{split}
            %%%%%%               P_p/m            %%%%%%
            &\dot{P}_\pm = 2 D_\pm \left[ \gamma^+ P_0  - \gamma^- P_\pm
            -\vec{\gamma}^- \cdot  \vec{S}_\pm \right], \\
            %%%%%%               Splus/minus          %%%%%%
            &\dot{\vec{S}}_\pm =  D_\pm \left[P_0 \vec{\gamma}^+   -\gamma^- \vec{S}_\pm - \tfrac{P_\pm}{2} \vec{\gamma}^-  \right] + 2 \vec{B}_\pm  \times  \vec{S}_\pm ,
        \end{split}
\end{equation}
complemented by $\dot{P}_0 = -\dot{P}_+ - \dot{P}_-$ due to probability conservation. Further simplifications apply if $\theta=0$, as the exchange field $\vec{B}_\pm$ reduces to $D_\pm \vec{\omega}_S$. Thus, in this limit,  $D_\pm$ factorizes in the equations of the spin variables. We are left with a single spin resonance with the condition given by $\vec{\omega}_S \cdot \left(\vec{n}^\text{L}_S-\vec{n}^\text{R}_S\right)=0$. Interestingly, the prefactors $D_\pm$ drop completely from the stationary solutions. They can simply be interpreted as scaling factors for the time evolution of the different channels. As such, they can not influence the stationary state, achieved in the infinite time limit.

In the case $\theta=\pi/2$, instead, the two spin variables are characterized by two independent resonant conditions $\vec{B}_\pm \cdot \left(\vec{n}^\text{L}_S-\vec{n}^\text{R}_S\right)=0$. The splitting of the resonances as a function of the angle and pseudospin polarization strength is highlighted in Fig.~\ref{fig:pol_angle_plot}. 

\begin{figure*}
    \includegraphics[width=500pt,draft=false]{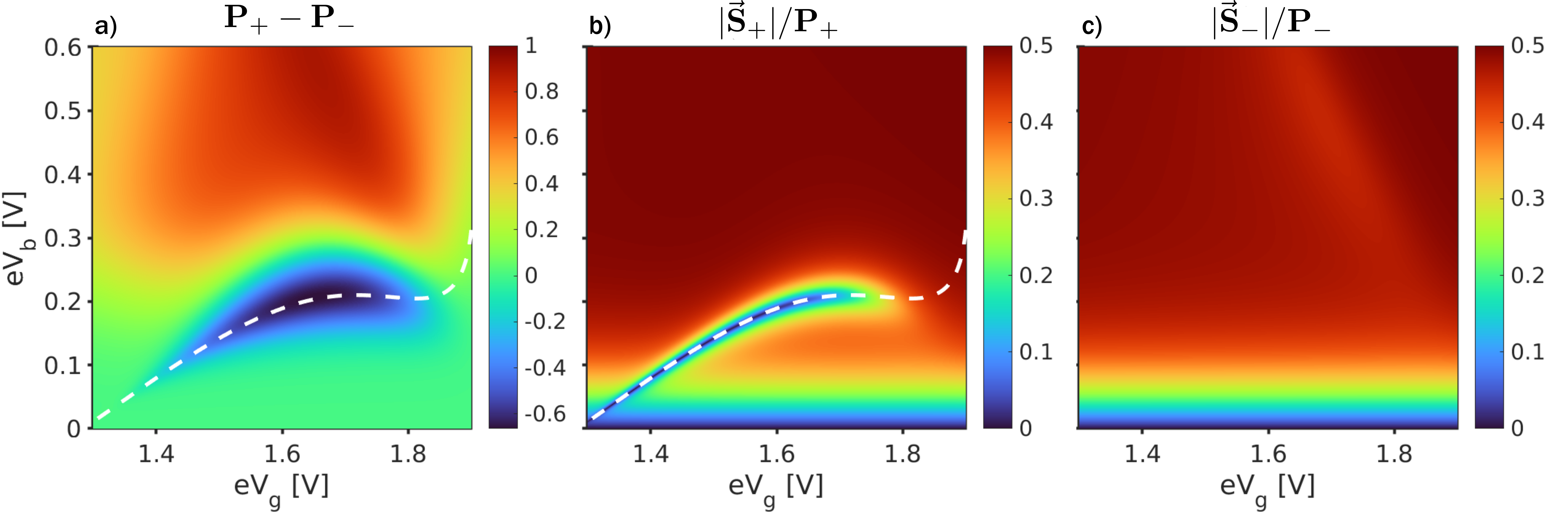} 
    \caption{Spin dephasing as the main mechanism behind spin resonances: a) The limiting case $\theta=\pi/2$ shows a splitting into the $\pm$ channels. The default situation is that the electrons occupy $P_+$ (red area) since the pumping is polarized in that direction. $P_-$ is prevailing on the resonance condition for the "+" channel (white dashed line) since there the "+" electrons can leave the spin valve blockade and thus only "-" electrons remain.  b) On the resonance, the spin coherence decreases faster than the respective population. c) Clear resonance condition for the "-" channel is outside of this $V_\text{g}$-$V_\text{b}$ window,  thus the coherence of this channel is maintained and yields a blockade. Parameters of Fig.~\ref{fig:overview_CD} except of $\theta=\pi/2$. \label{fig:limiting_cases} }
\end{figure*}

In Fig.~\ref{fig:limiting_cases}, we further analyze the spin dynamics underlying such resonances.  The observable $P_+-P_- \equiv 2\vec{T}\cdot \et$ shows a very strong transfer of probability from the "+" towards the "-" channel in the vicinity of the $\vec{S}_+$ spin resonance (highlighted by the white dashed line). A comparison with Fig.~\ref{fig:limiting_cases}\,b) indicates, moreover, how the spin dephasing is at the origin of the population transfer. The fast precession opens the "+" spin channel, and the \emph{average} spin amplitude (perpendicular to the exchange field) drops even faster than the corresponding population. Spin accumulation for the slow precessing "-" channel completes the picture. This observation contrasts, though, with the picture of coherent rotation as unblocking mechanism, as the latter would conserve the rotated spin length, or at least, the ratio between the spin and the corresponding population.

The understanding of the limiting cases allows us to infer a similar dynamics for $\theta=1.5\approx\pi/2$. Fig.~\ref{fig:pol_angle_plot}\,b) shows how the resonances predicted for $\theta = \pi/2$ closely follow two of the actual resonances. The other two resonances of this plot can be rationalized, instead, by the semi-analytical ansatz of Eq.~\eqref{eq:ffunction} as a delicate interplay of the pumping vector and the involved magnetic fields. The elements $\ey \cdot \vec{T}$ and $\eperp \cdot \vec{T}$ feed into the spin channels and cause, there, an accumulation of spin components which are eventually not blocked. 

Remarkably, in the areas where both unblocking conditions were meet simultaneously, i.e.\ in Fig.~\ref{fig:current_comparison}  around the anti-crossing of $V_\text{g}\approx1.6$ and $ V_\text{b}\approx0.2$, a near to perfect lifting of the spin blockade is reached. The current closely approaches the one that would be obtained for normal leads, in the complete absence of spin valve.

%%%%%%%%%%%%%%%%%%%%%%%%%%%%%%%%%%%%%%%%%%
%%%%%%%%%%%%%%%%%%%%%%%%%%%%%%%%%%%%%%%%%%

\section{Entanglement of spin and pseudospin}
\label{sec:entanglement}

\begin{figure}
    \includegraphics[width=240pt,draft=false]{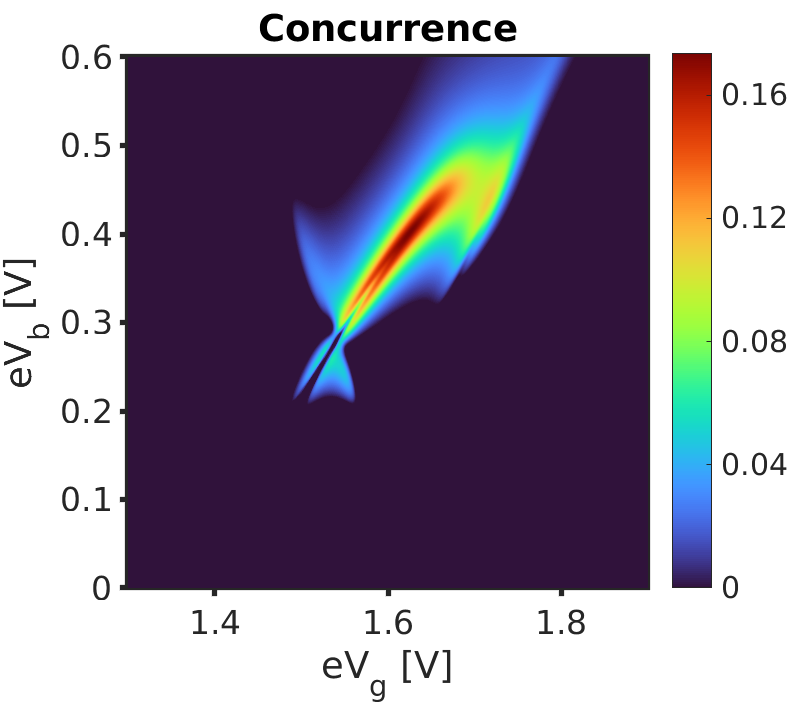} 
    \caption{Concurrence $C$ in dependence of gate and bias voltage: Remarkably, only in a limited area entanglement between spin and pseudospin can be observed. These features are determined by the correlator vectors $\vec{\Lambda}_y$ and $\vec{\Lambda}_\perp$. Same parameters as in Fig.~\ref{fig:overview_CD}. \label{fig:concurrence}}
\end{figure}

The interaction between the spin and the pseudospin, discussed in the previous section and triggered by an intermediate pseudospin polarization angle, yields not only correlation but also entanglement between the two degrees of freedom. As a measure of the phenomenon, we choose the \emph{concurrence} (see Fig.~\ref{fig:concurrence}) which, together with the closely related \emph{entanglement of formation}, quantifies the degree of quantum entanglement of a system \cite{Wootters1998}.
In particular, for a bipartite system, the entanglement of formation is calculated as 
\begin{equation}
    \mathcal{E}(C) = h\left(\frac{1+\sqrt{1-C^2}}{2}\right),
\end{equation}
where $h(x)=-x \log_2 x -(1-x)\log_2(1-x)$ is the Shannon entropy function \cite{Hill1997} and $C$ is the concurrence. The entanglement of formation $\mathcal{E}$ ranges from 0 to 1 and is a monotonically increasing function of the concurrence. Since the concurrence $C$ also ranges from 0 to 1, it can also be seen as standalone measurement of entanglement. 

The concurrence is obtained by different, though compatible, formulas depending on the state of the system. For a pure state $\ket{\Psi}$, $C$ is given as 
\begin{equation}
    C(\Psi)=|\langle \Psi|\tilde{\Psi}\rangle|,
\end{equation}
where $|\tilde{\Psi}\rangle = \sigma^y \otimes \sigma^y \ket{\Psi^*}$ with $^*$ the complex conjugation which, together with the two $\sigma^y$'s represents the spin flip operation for each of the two involved degrees of freedom. For the one-particle sector of our DQD, we can express this combined spin flip with the help of the spin and pseudospin operators:  $|\tilde{\Psi}\rangle = 4 \hat{S}_y \hat{T}_y \ket{\Psi^*}$. 

Notice that, with our choice, we do \emph{not} measure the concurrence between the spin on each dot, but rather the one between the \emph{spin} and the \emph{pseudospin} of the full DQD. On this respect, the state  $(\ket{\uparrow,0}-\ket{0,\downarrow})/\sqrt{2}$ is an example of maximal entanglement. It yields concurrence 1 because a simultaneous pseudospin and spin flip, up to a sign, does not alter the state. In contrast,  $(\ket{\uparrow,0}-\ket{0,\uparrow})/\sqrt{2}$ gives zero concurrence, since the simultaneous flip of spin and pseudospin leads to a state orthogonal to the original one.

We are interested, though, in the concurrence for a generic state of our bipartite pseudospin-spin DQD, described by the one-particle component $\hat{\rho}_\text{red,1}$ of the reduced density matrix. Following \cite{Wootters1998}, we thus calculate the concurrence as
\begin{equation}
\label{eq:}
    C\left(\hat{ \rho}_\text{red,1} \right)=\text{max}(0,\lambda_1-\lambda_2-\lambda_3-\lambda_4),
\end{equation}
where the $\lambda_i$'s are the square roots of the eigenvalues, in decreasing order, of the non-Hermitian matrix $\hat{\rho}_\text{red,1} \tilde{\rho}_\text{red,1} $. Analogously to $|\tilde{\Psi}\rangle$, we define the pseudospin- and spin-flipped state as $\tilde{\rho}_\text{red,1}= (\sigma^y \otimes\sigma^y ) \hat{\rho}_\text{red,1}^* (\sigma^y \otimes\sigma^y )$.  

In Fig.~\ref{fig:concurrence}, the concurrence of our system is displayed in dependence of bias and gate voltages. One appreciates how quantum mechanical entanglement of spin and pseudospin is only present on the resonances ("cross"-shaped structures) which are not captured by the limiting case of the independent "+" and "-" spin channels. Consequently, the finite values of the concurrence closely correlate to the $\ey$ and $\eperp$ components of $\tvec$ shown in  Fig.~\ref{fig:pseudospin}a) and c). The mediator of the entanglement between the spin and pseudospin in the DQD is the synthetic spin-orbit interaction induced by the electronic fluctuations. 

With the concurrence, we can quantify and compare the degree of entanglement with respect to other systems or polarization configurations. In graphene, for example, entanglement between spin and sublattice pseudospin leads to the formation of states which violate the Bell inequality \cite{deMoraes2020}. The latter should be detectable via Cooper pair splitting experiments. The time-varying concurrence generated in graphene by the intrinsic spin-orbit interaction, ranges in their calculation between 0.5 and 0.6. Beyond its relevance for fundamental physics, the study of entanglement \cite{Magazzu2018} is crucial for the development of current quantum technologies. In this spirit, the discussed electrical manipulation of quantum entanglement represents a new interesting path for the implementation of qubit operations in DQD.    

%%%%%%%%%%%%%%%%%%%%%%%%%%%%%%%%%%%%%%%%%%
%%%%%%%%%%%%%%%%%%%%%%%%%%%%%%%%%%%%%%%%%%

\section{Experimental realization}
\label{sec:experimental_realization}
To our knowledge, spin/pseudospin resonances, as the ones outlined in this publication, have - up to now - not been realized experimentally. However, we are convinced that, though very challenging, these experiments are feasible. Pseudospin resonances bear the advantage that a huge variety of systems exhibit the necessary twofold degeneracy in their valley/orbital degree of freedom, and moreover, that the necessary high pseudospin polarizations of the leads have been already implemented in interference experiments \cite{Donarini2019}.

In general, suitable candidates for detection of these resonances are quantum dots realized in carbon nanotubes \cite{Donarini2019,Aurich2010}, in semiconductors \cite{Bordoloi2020} or in molecules within a STM setup \cite{Zheng2017}. The quest is to combine the two main prerequisites which are already individually achieved in experiments: On the one side, a valve configuration \cite{Bordoloi2020} and on the other side off-diagonal tunneling rate matrices, i.e.\ coherent tunneling \cite{Karlstroem2011,Donarini2019}. For example, in \cite{Rohrmeier2021} we elaborate theoretically a microscopic model of DQD where such off-diagonal tunneling rate matrices are obtained manipulating the distance between the quantum dots and with respect to the leads.

\section{Conclusions}
\label{sec:conclusion}

The transport characteristics of interacting systems with a degenerate many-body spectrum are prone to exhibit interference effects \cite{Braun2004,Begemann2008,Karlstroem2011,Hell2013} already in  the sequential tunneling regime. Interference appears whenever the single-particle tunneling matrices of the leads (see Eq.~\eqref{eq:Gamma_def}) cannot be diagonalized simultaneously. In other terms, whenever it is not possible to identify parallel transport channels running between the source and the drain lead. 

In this work, we analyzed an interacting DQD weakly coupled to ferromagnetic leads in almost antiparallel spin valve configuration. This set up naturally ensures interference between the \emph{spin} transport channels. Moreover, we choose a tunneling coupling with parallel pseudospin polarization, which, naively, should correspond to independent pseudospin channels.

On the other hand, the tendency of the electrons to avoid each other due to the Coulomb interaction induces pseudospin anisotropy on the DQD, thus defining a pseudospin hard axis. It is the angle $\theta$ between this axis and the polarization direction of the leads to control the mixing of the pseudospin channels. 

For $\theta = 0$, the stationary pseudospin is completely quenched and the dynamics reduces to the one of a quantum dot spin valve \cite{Hell2015}. In the case of $\theta = \pi/2$, instead, we can identify two different spin variables, $\vec{S}_+$ and $\vec{S}_-$ associated with opposite pseudospin directions and showing independent dynamics. Thus, the pseudospin reduces itself to a single component, the one parallel to the lead polarization, which measures the imbalance $P_+ - P_-$ between the populations of the two spin channels. Finally, for any other intermediate angle, the spin and the pseudospin are correlated, with the stationary pseudospin changing strength and direction as a function of the bias and gate voltage applied to the system.   

We focused on the angle $\theta = 1.5 \approx \pi/2$. Here, the signatures of the intertwined spin and pseudospin dynamics are current resonances emerging inside the one-particle Coulomb diamond. Besides the spin resonances closely related to the ones of the limiting case with $\theta = \pi/2$, we identify a cross-shaped feature which can only be understood in terms of spin-pseudospin correlations. 

In general, all the observed current resonances result from the lifting of the spin blockade induced by the spin valve configuration. The exchange fields induce a fast precession of the spin variables, which results in spin dephasing. Thus, the electrons can again tunnel towards the drain, despite its high spin polarization. In particular, the direction of the exchange fields controls the efficiency of the dephasing and thus the position of the spin resonances within the Coulomb diamond. The cross-shaped resonance, instead, stems from the interplay of spin and pseudospin and their mutual influence in their pumping dynamics, where also the correlation vectors $\vec{\Lambda}_y$ and $\vec{\Lambda}_\perp$ are involved. Ultimately, we could show that, in the vicinity of the cross-shaped resonance, spin and pseudospin are not only correlated, but also entangled. To this end, the calculation of the concurrence gives a figure of merit for the effect.    
A fundamental issue addressed in this study is the emergence of spin-pseudospin correlation and entanglement, despite the factorized form of the tunneling matrices. Moreover, the different nature of the current resonances observed in the one-particle Coulomb diamond shows how to address different transport channels and stir the dynamics of different degrees of freedom of an interacting system solely by electrical means, i.e.\  the bias or the gate voltages across the nanojunction. 

Systems with larger ($N > 2$) level degeneracy exhibit a coherent dynamics involving a rapidly increasing number of degrees of freedom. Together with their fast increasing complexity, though, they also offer more control nobs. Modulating the tunneling amplitudes between a multilevel system and the leads induces variations of the exchange fields arising from electronic fluctuation. Ultimately, the results presented here indicate, in principle, how to achieve in a single device, an all electronic control of the precession dynamics for several entangled degrees of freedom, a very desirable feature for the current quest of a scalable quantum information technology.

%%%%%%%%%%%%%%%%%%%%%%%%%%%%%%%%%%%%%%%%%%
\section*{Acknowledgments}
%%%%%%%%%%%%%%%%%%%%%%%%%%%%%%%%%%%%%%%%%%
We thank M.~Grifoni and L.~Magazzù for fruitful discussions. The authors acknowledge moreover the financial support from the Elitenetzwerk Bayern via the IGK Topological Insulators and the Deutsche Forschungsgemeinschaft via the SFB 1277 (subprojects B02 and B04). 

%%%%%%%%%%%%%%%%%%%%%%%%%%%%%%%%%%%%%%%%%%
%\section{Appendix}
%%%%%%%%%%%%%%%%%%%%%%%%%%%%%%%%%%%%%%%%%%

\bibliography{bibliography}

\end{document}